\documentclass[twocolumn]{article}
\usepackage[margin=1in]{geometry}

\usepackage{tikz}
\usepackage{xcolor}

\usepackage{multirow}
\usepackage{multicol}

\usepackage{pifont}
\usepackage{varioref}

\usepackage{eucal}
\usepackage{subfig}

\usepackage{rotating}
\usepackage{colortbl} 

\usepackage{amsmath,amssymb,amsfonts}
\usepackage{enumerate}
\usepackage{tablefootnote}
\usepackage{threeparttable}
\usepackage{soul}
\usepackage{booktabs}
\usepackage{algorithm,algpseudocode}

\usepackage[font={footnotesize}]{caption}

\usepackage{enumitem,kantlipsum}
\usepackage{array}
\usepackage{tablefootnote}
\usepackage{threeparttable}
\usepackage[toc,title,page]{appendix}
\usepackage[hyphens]{url}
\usepackage[colorlinks=true,urlcolor=blue]{hyperref}
\usepackage{textcomp}
\usepackage{bmpsize}
\usepackage{stfloats}

\newcolumntype{P}[1]{>{\centering\arraybackslash}p{#1}}

\newcommand{\bluetext}[1]{\textcolor{black}{#1}}

\algrenewcommand\algorithmicrequire{\textbf{Input:}}
\algrenewcommand\algorithmicensure{\textbf{Output:}}

\begin{document}

\title{Privacy and Security of Women's Reproductive Health Apps in a Changing Legal Landscape}

  \author{
  Shalini Saini\\
  \texttt{Texas A\&M University}
  \and
  Nitesh Saxena\\
  \texttt{Texas A\&M University}
}
\date{}
\maketitle
\begin{abstract}

FemTech, a rising trend in mobile apps, empowers women to digitally manage their health and family planning. However, privacy and security vulnerabilities in period-tracking and fertility-monitoring apps present significant risks, such as unintended pregnancies and legal consequences. Our approach involves manual observations of privacy policies and app permissions, along with dynamic and static analysis using multiple evaluation frameworks. Our research reveals that many of these apps gather personally identifiable information (PII) and sensitive healthcare data. Moreover, around 85\% of the app privacy policies examined lack explicit mention of security measures, despite the sensitivity of health data. Furthermore, our analysis identifies that 61\% of the code vulnerabilities found in the apps are classified under the top-ten Open Web Application Security Project (OWASP) vulnerabilities. 

Our research emphasizes the significance of tackling the privacy and security vulnerabilities present in period-tracking and fertility-monitoring mobile apps. By highlighting these crucial risks, we aim to initiate a vital discussion and advocate for increased accountability and transparency of digital tools for women's health. We encourage the industry to prioritize user privacy and security, ultimately promoting a safer and more secure environment for women's health management.

\end{abstract}

\label{sec:intro}
\section{Introduction}

Digital reproductive health management meets modern medicine needs for better accessibility and continuous monitoring. Female Technology (Femtech), particularly period-tracking and fertility monitoring apps, addresses women's unique health needs \cite{femtech}. The Femtech market has surged tenfold in the past decade and is projected to reach a value of \$50 billion by 2025 \cite{femtech-statista}. However, in the absence of enforced regulations, these apps are susceptible to privacy and security attacks.

Smartphones have revolutionized healthcare by facilitating remote recording, sharing, and access to medical assistance, particularly during the pandemic. In North America, 88\% of the population (329 million mobile users) and 75\% of networked devices/connections will be connected via Wi-Fi \cite{cisco23networkstates}. Mobile technologies can enhance sexual and reproductive health by offering education, supporting vulnerable populations, and optimizing reproductive health through features like fertility tracking apps, contraceptive counseling, and birth control reminders \cite{chandler2020promoting, zvarikova2022menstrual, greene2021acceptability}. However, increased network connectivity poses a risk of malicious attackers accessing users' private information. Reproductive health apps, leveraging users' needs and interests, may encourage the sharing of private data over the network, which can be exploited for surveillance capitalism \cite{zuboff2019surveillance, ford2021hormonal}.

\begin{figure*}[]
  \centering
  \includegraphics[width=.8\linewidth]{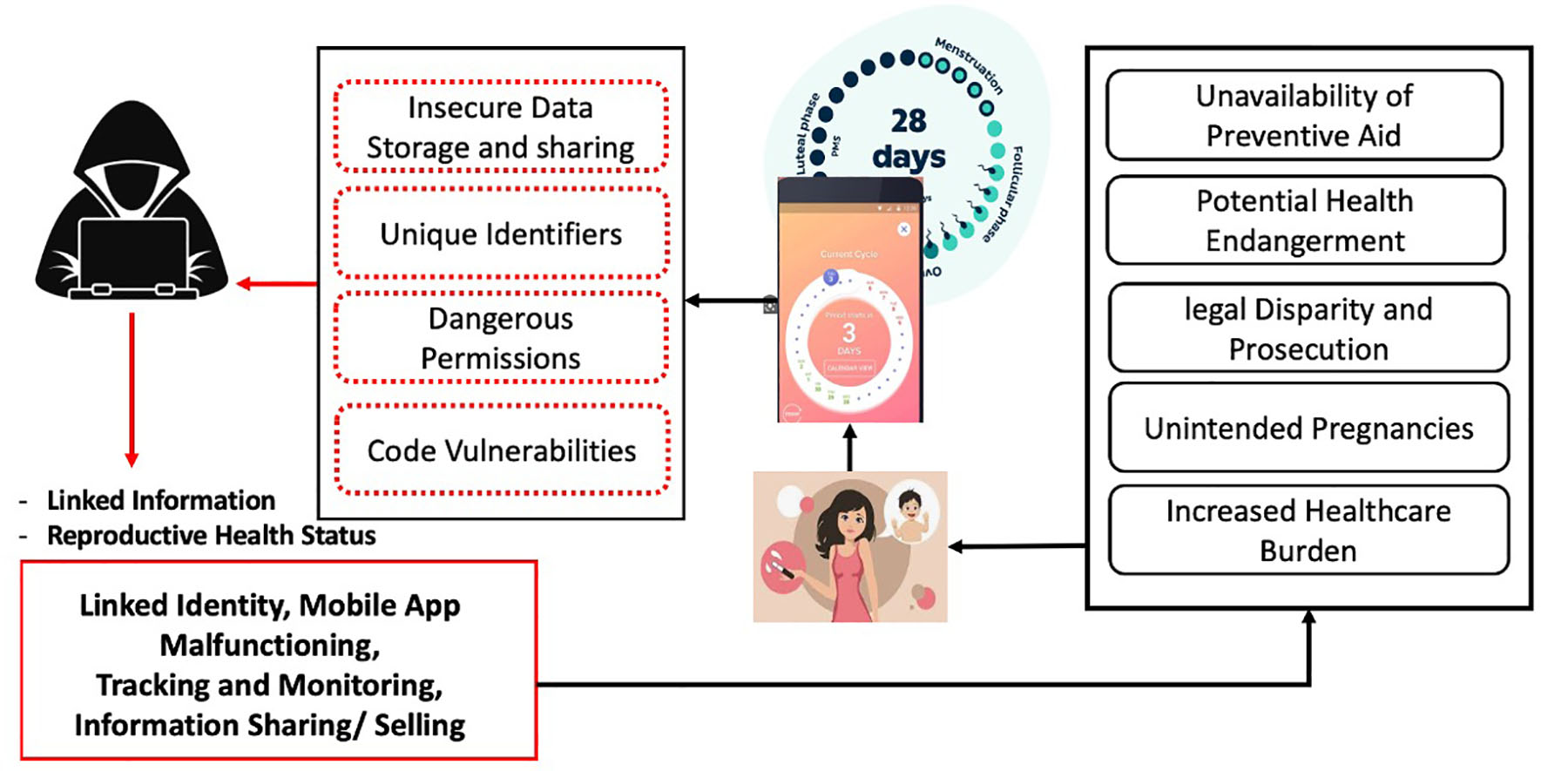}
	\caption{Privacy and Security Threats of Vulnerabilities of Reproductive Mobile Apps to Women's Health} 
	\label{fig:pt-threat}
\end{figure*}

Mobile apps face challenges in validating integrity, reliability, effectiveness, privacy, security, and accountability, unlike regulated medical devices. The FDA approved \textit{Natural Cycles} in 2018 as the first mobile medical application for contraception \cite{nc_fda}. However, it has a typical use failure rate of up to 8.3\% \cite{bull2019real}. If an app fails to prevent pregnancy, it can result in lawsuits and financial claims \cite{NaturalCycles-failure}. Lack of regulation and compliance verification allows app developers to exploit user data for maximum monetization, disregarding privacy and security. Certain permissions, default settings, bundled permissions, and data sharing can lead to such exploitation \cite{consumerreports2020}. Figure \ref{fig:pt-threat} illustrates the potential threats and negative impacts of exploiting vulnerabilities in reproductive health apps, emphasizing the harm caused by malicious manipulation of these tools. Our study makes several contributions including:
\begin{itemize}
	\item Our work verifies that majority of studied reproductive health apps collect Personal Identification Information (PII), geographical data, and sensitive health data.
	\item The study results reveal that dangerous permissions and vulnerabilities can grant access to restricted app activities and data, which could potentially facilitate privacy and security attacks.
	\item The study highlights that in the post Roe v. Wade era, period-tracking and fertility-monitoring apps exhibit vulnerabilities that significantly increase the risk to women's privacy and security
	\item \bluetext{The study findings call for collaboration among health professionals, developers, and policy-makers to develop secure and trustworthy digital health management apps for women's health.}
\end{itemize} 

With Roe v. Wade overturned, period-tracking and pregnancy managing mobile apps may be vulnerable to exploitation for tracking and identifying vulnerable populations for legal prosecution \cite{harris2022navigating, ms-magazine-postroe-pt-privacy2022}. The vulnerability of users' sensitive health data in reproductive healthcare apps is a significant concern due to law enforcement's ability to access data from third-party servers or the user's device through subpoenas. Regulatory standards are necessary to ensure the protection of users of reproductive healthcare mobile apps, despite the guidelines issued by the U.S. Department of Health and Human Services (HHS). The Federal Trade Commission (FTC) safeguards sensitive health data, even in cases not regulated by The Health Insurance Portability and Accountability Act (HIPAA) \cite{hhs-hipaa-mobile-postroe, ftc-health-rule}. The recent case of a period-tracking app being barred by the Federal Trade Commission (FTC) for selling personal data to advertisers is just one example of the exploitation of unregulated reproductive healthcare apps \cite{ftc-bars-app-May2023}.

In areas where abortions are illegal, manipulated apps not only increase the risk of unintended pregnancies, leading to health and social crises, but also expose individuals to potential legal repercussions. The study highlights significant privacy and security vulnerabilities in even the most popular period-tracking and fertility-monitoring apps, despite their well-established market presence. These risks are particularly worrisome for women facing disadvantages, including limited access to essential healthcare and preventive interventions \cite{chandler2020promoting}. Therefore, it is crucial to regulate secure FemTech apps to protect the women's reproductive health rights.

\label{sec:background}
\section{Background}

The 2020 CDC Abortion Surveillance Report shows that 615,911 abortions were reported, with an abortion rate of 11.2 per 1,000 women aged 15-44 years \cite{cdc-abortion-2020}. Unintended teen pregnancies pose a social, medical, and financial burden. Teen pregnancy rate dropped 64\% between 1991 and 2015, resulting in \$4.4 billion in public savings in 2015 alone \cite{ptd-2018, cdc-teen-preg-2019}. However, 30\% of women aged 13-44 in the United States are in need of publicly funded contraception, and 98\% lack reasonable access to these resources, particularly vulnerable populations such as young women in foster care \cite{cdc-teen-preg-2019, ptd-2018}.
\subsection{Health Risks of Legal Disparity}
The right to abortion is an essential liberty tied to basic rights such as family matters and bodily autonomy \cite{roe-v-wade}. Unsafe abortion practices pose grave risks, including potential fatalities. However, accessing safe and legal abortion services can be difficult, especially for vulnerable populations. The COVID-19 pandemic has compelled healthcare providers to adopt telemedicine, but this shift raises concerns regarding privacy and security. Furthermore, the overturning of Roe v. Wade has sparked fears that individuals' menstrual cycle data obtained through apps could be exploited for legal prosecution purposes \cite{harris2022navigating, ms-magazine-postroe-pt-privacy2022}. Period-tracking mobile apps may falsely claim to protect privacy and data security, and unauthorized access to healthcare-related data can be exploited. More evidence-based studies are needed to build trust in their clinical utility, efficacy, and safety. There are known biases and trends that can result in targeted attacks for the purpose of monetizing these predictions, and to gain behavioral control \cite{zuboff2019surveillance}.

\subsection{Privacy, Security, and Legal Risks of Women's Health Apps}

Reproductive health management mobile apps can assist women to be aware on fertile cycle and making conscious decision about the pregnancy, but collecting and misusing the app data can pose legal threats. There is a possibility of biased legal actions towards minorities and widening the inequality gap \cite{harris2022navigating, stevenson2021pregnancy}. 

Period-tracking apps have raised concerns about user data security and privacy, particularly after the overturning of Roe v. Wade, which could potentially criminalize users \cite{campanella2022menstrual}. Leading apps like Flo and Clue have issued statements to protect user privacy. Flo also introduced an anonymous mode \cite{flo-anonymous, flo-anonymous-whitepaper}. App developers face challenges in assuring users while complying with legal limitations. 

Digital healthcare can reduce the burden of lower status healthcare workers, including low-income women of color for managing healthcare and family planning \cite{scales2020s}. However, changes in medicine practices due to the COVID-19 pandemic and restrictions on abortion clinics have led to a significant increase in medication abortions, potentially putting app users at risk of prosecution for non-compliance with abortion laws if their identity is linked through vulnerabilities in the app \cite{allsworth2022telemedicine}.

It is interesting to know how non-technical issues may impact the privacy and security vulnerabilities of menstruation and fertility apps. For example, Aljedaani et al. showed that 63\% of mobile health apps lack security because of absent security guidelines and regulations for developing secure  mobile health apps \cite{aljedaani2021challenges}. Moreover, as per Earle et al., the limited involvement of health professionals or users in the design, development, and deployment of period-tracking and fertility apps may restrict developers' ability to fully grasp the sensitivity of reproductive health data \cite{earle2021use}.

\label{sec:related work}
\section{Related Work}
Considering the sensitive nature of reproductive data, Shipp and Blasco conducted an analysis of 30 mobile apps that focus on fitness and health, specifically those tracking period data. Their primary focus was on analyzing the privacy policies of these apps, mostly manually, and highlighting the lack of specific information regarding the collection of period data \cite{shipp2020private}. In comparison, our work includes a comprehensive coverage of privacy policies, code vulnerabilities, dangerous permissions, and runtime vulnerabilities in women's health apps, with a total of approximately 144 million downloads. It is crucial to emphasize that no single tool can detect all vulnerabilities effectively.

While there are efforts to discuss the practical benefits of digital health management for reproductive health, Lupton highlights significant ethical and privacy implications of self-tracking through reproductive activity tracking apps and the data they produce \cite{lupton2015quantified}. Another work by Epstein et al. collected and analyzed data from multiple sources to understand menstrual cycle tracking practices by reviewing 2,000 app reviews, surveying 687 participants, and conducting follow-up interviews with 12 individuals. Findings reveal that women track their cycles for various reasons, including period prediction and healthcare discussions. Six tracking methods were identified, including technology use and physiological awareness. However, accuracy issues were found with apps and calendars. Gender and sexual minorities felt excluded, and life stages like young adulthood and menopause were overlooked \cite{epstein2017examining}.

There is a challenge to ensure that healthcare apps, which curate sensitive health data, undergo rigorous evaluation and meet accuracy standards \cite{curchoe2020smartphone}. A 2020 study by LaMalva and Schmeelk shows that MobSF analysis identified at least one OWASP-defined security violation in 43.6\% of studied healthcare android apps \cite{lamalva2020mobsf}. Alfawzan et al. found that all women's health apps, including reproductive health apps, allowed behavioral tracking and had poor privacy practices \cite{alfawzan2022privacy}. 

Iyawa et al. found that mobile apps for self-management during pregnancy show positive impacts. however, mobile apps can be further studied for the identification of sexually transmitted infections, early warning signs of complexities during pregnancy and miscarriage \cite{iyawa2021mobile}. A 2019 study by Grundy et al. showed that 79\% of Google Play medicine related health apps regularly shared user data, and short of maintaining  transparency with the app users. Studied apps demonstrate privacy leaks inferred with sharing sensitive information to remote server. User's privacy is threatened by apps receiving sensitive user data, including IP addresses and GeoIP \cite{Grundyl920}. Bull et al. analyzed data from the Natural Cycles app and found that tracking physiological parameters such as basal body temperature, rather than just cycle length, is critical to clinically identifying the fertile period \cite{bull2019real}. A study found that the only FDA-approved Natural Cycles app, a highly effective hormone-free contraceptive method compared to traditional methods, has a failure rate of 8.3\% and can produce unintended outcomes \cite{scherwitzl2017perfect}. The review of European apps highlights with evidence suggesting that some apps are useful for women who do not want to rely on other methods of contraception. However, not all apps accurately predict the fertile window, leading to potential risks for women relying on them for pregnancy prevention. The fluid usage of fertility apps over time increases the likelihood of unintended pregnancies \cite{earle2020use}.

Reardon et al. conducted an investigation into information leakage in hundreds of thousands of Android apps, primarily using dynamic analysis \cite{reardon201950}. In contrast to their research on general apps, our study focuses on domain-specific Femtech apps and examines risks within their unique contexts. We use multiple tools to cross-verify findings from static and dynamic analysis methods, in addition to analyzing privacy policies. Our comprehensive approach helps us identify more vulnerabilities collectively, which is particularly significant given the evolving legal complexities surrounding abortion.

\label{sec:Study Design}
\section{Research Objectives}
\bluetext{Our research study is focused on a meticulous examination of the most widely accessible and highly popular Android apps designed for tracking menstrual cycles and ovulation in order to provide a thorough understanding of privacy and security vulnerabilities.} Our analysis of these apps' privacy policies, dangerous permissions, exploitable attack surface, and code flaws revealed that vulnerable apps pose a substantial risk to users' privacy and security. Considering the Post Roe v. Wade abortion rule in the United States, these risks may have legal implications. \bluetext{Our overarching aim is to comprehensively identify and evaluate the privacy and security vulnerabilities inherent in reproductive health tracking apps. This includes meticulous scrutiny of app permissions, as we acknowledge that while certain permissions are essential for app functionalities, they can still be exploited, posing potential risks to privacy and security.} Our research objectives are as following:

\begin{itemize}
\item Analyze the availability of privacy policies and the information transparency.
\item Analyzing the vulnerabilities associated with identifiable information collection and sharing.
\item Identifying the exploitable App permissions, dangerous to manipulate app behavior and mishandle sensitive health data.
\item Identifying the app threat surface which can cause the unintended app behavior and undesirable outcomes.
	
\end{itemize}

\begin{table*}[]
\centering
\caption{Fertility, Ovulation, and Period-Tracking Mobile Apps-Downloads (M- Million; K-Thousand ; +- More than), 
Ratings (out of 5) and \# of Reviewers}
\label{tab:apps-basic-data}
\begin{tabular}{|p{5cm}|p{2cm}|P{4cm}|P{3cm}|P{4cm}|P{1.5cm}|}

\hline
\multirow{2}{*}{App   Name}       & \multirow{2}{*}{Downloads} & \multicolumn{2}{l|}{iOS}                & \multicolumn{2}{l|}{Android}          \\ \cline{3-6} 
                                  &                            & \multicolumn{1}{l|}{Rating} & Reviews   & \multicolumn{1}{l|}{Rating} & Reviews \\ \hline
Always You:   Period Tracker      & 100K+                      & \multicolumn{1}{l|}{4}      & 153       & \multicolumn{1}{l|}{3.4}    & 341     \\ \hline
Birth Control   - Natural Cycles  & 1M+                        & \multicolumn{1}{l|}{4.8}    & 14,771    & \multicolumn{1}{l|}{4.7}    & 19.8K   \\ \hline
Clover - Safe   Period Tracker    & 1M+                        & \multicolumn{1}{l|}{4.7}    & 6,597     & \multicolumn{1}{l|}{4.5}    & 157K    \\ \hline
Clue Period   \& Cycle Tracker    & 10M+                       & \multicolumn{1}{l|}{4.8}    & 340,427   & \multicolumn{1}{l|}{4.3}    & 1.17M   \\ \hline
Eve Period   Tracker              & 1M+                        & \multicolumn{1}{l|}{4.7}    & 107,014   & \multicolumn{1}{l|}{4.4}    & 26.5K   \\ \hline
Fertility   Friend- FF App        & 1M+                        & \multicolumn{1}{l|}{4.8}    & 6,958     & \multicolumn{1}{l|}{4.8}    & 18K     \\ \hline
Flo                               & 100M+                      & \multicolumn{1}{l|}{4.8}    & 1,051,482 & \multicolumn{1}{l|}{4.6}    & 3M      \\ \hline
Glow                              & 1M+                        & \multicolumn{1}{l|}{4.7}    & 65,426    & \multicolumn{1}{l|}{4.3}    & 70.6K   \\ \hline
Kindara:   Fertility Tracker      & 100K+                      & \multicolumn{1}{l|}{4.7}    & 8,837     & \multicolumn{1}{l|}{3.2}    & 2.02K   \\ \hline
Luna                              & 5K+                        & \multicolumn{1}{l|}{4.3}    & 9         & \multicolumn{1}{l|}{3.7}    & 61      \\ \hline
MagicGirl/Teen   Period Tracker   & 500K+                      & \multicolumn{1}{l|}{4.6}    & 1,083     & \multicolumn{1}{l|}{4.6}    & 7.36K   \\ \hline
Maya                              & 5M+                        & \multicolumn{1}{l|}{4.8}    & 2,437     & \multicolumn{1}{l|}{4.7}    & 241K    \\ \hline
My Calendar                       & 10M+                       & \multicolumn{1}{l|}{4.8}    & 33,234    & \multicolumn{1}{l|}{4.8}    & 420K    \\ \hline
MyDays X                          & 1M+                        & \multicolumn{1}{l|}{3.6}    & 515       & \multicolumn{1}{l|}{4.1}    & 46.3K   \\ \hline
Ovia Fertility   \& Cycle Tracker & 1M+                        & \multicolumn{1}{l|}{4.8}    & 66,500    & \multicolumn{1}{l|}{4.6}    & 75K     \\ \hline
Period Diary                      & 500K+                      & \multicolumn{1}{l|}{4.7}    & 66,984    & \multicolumn{1}{l|}{2.8}    & 9.29K   \\ \hline
Period Tracker                    & 10M+                       & \multicolumn{1}{l|}{4.8}    & 61,374    & \multicolumn{1}{l|}{4.6}    & 360K    \\ \hline
Premom   Ovulation Tracker        & 1M+                        & \multicolumn{1}{l|}{4.7}    & 18,869    & \multicolumn{1}{l|}{4.1}    & 10.4K   \\ \hline
Spot On Period   Tracker          & 500K+                      & \multicolumn{1}{l|}{4.3}    & 15,211    & \multicolumn{1}{l|}{4.2}    & 7.33K   \\ \hline
Stardust   Period Tracker         & 50K+                       & \multicolumn{1}{l|}{4.2}    & 12,645    & \multicolumn{1}{l|}{4.4}    & 2.31K   \\ \hline
\end{tabular}
\end{table*}

\label{sec:Methodology}
\section{Methodology}
We analyzed 20 period-tracking and fertility-monitoring mobile apps available for iOS and Android devices. We compiled a list of these apps by verifying web search and app store details, considering criteria such as availability in English, free download, high ratings, and a substantial number of reviewers (indicating credibility and user trust). These apps have been recommended as top choices for managing women's health digitally by various online resources \cite{top-apps, top-apps-2}. Table \ref{tab:apps-basic-data} presents the details of the considered period-tracking and fertility-monitoring apps in this work.

\subsection{Mobile App: Components}

A mobile app consists of a User Interface (UI), app components, functionalities, and permissions. The UI is represented by an Activity with multiple screens, while a non-UI component called Service performs background operations. The Content Provider enables data sharing within the app or with other apps, and the Broadcast Receiver verifies the legitimacy of intents from authorized sources \cite{cwe}. App permissions grant access to device features and data resources required for intended functionality. Android categorizes \textit{Dangerous} permissions that can provide apps with additional access to restricted data or actions. Attackers can exploit such permissions to access sensitive information without the user's consent \cite{reardon201950, androidPermissions}.

\subsection{Vulnerability Analysis Methods}

\subsubsection{Static Analysis}
Static Analysis is performed using Android Studio 3.6.1. to analyze the source code of an application to identifying the code vulnerabilities without running the code. We acquired the latest APK file for all studied apps through an Android mobile app \textit{APK Extractor}. Extracted .APK files are decompiled with an online APK decompiler\footnote{\href{http://www.javadecompilers.com/}{\textit{APK decompiler}}.}. 

\subsubsection{Dynamic Analysis}
Android dynamic testing tool \textit{Drozer 2.3.4} \cite{drozer} is used for dynamic analysis for finding the attack surface and the app permissions. Drozer is a well-known, open-source security assessment tool for Android apps \cite{drozer}. Test mobile device has a Drozer client installed, and Drozer commands were executed on Windows 10 machine. 

\subsubsection{MobSF Analysis}
Mobile Security Framework (MobSF) \footnote{MobSF https://mobsf.live/} is an automated, all-in-one mobile application (Android/iOS/Windows) pen-testing, malware analysis and security assessment framework. MobSF analyzes an individual .apk file uploaded to a Web interface. Mobsf generates a report containing information regarding the app's security and privacy vulnerabilities. The comprehensive report generated by MobSF categorizes vulnerabilities into different categories, including network security, permissions, code analysis, and more. This report offers valuable information to assess the overall security status of the app, providing risk ratings ranging from low to medium and high. 

\subsection{Vulnerability Analysis Categories }
We present the privacy and security vulnerabilities in four major categories of Privacy Policy Analysis, Permissions Analysis, Attack Surface Analysis, and Code Analysis.
 
\subsubsection{Privacy Policy Analysis}
We manually analyze all app policies by extracting details related to the information being collected, the sensitivity of the collected information, data handling and sharing policies. Additionally, we gather information on data retention policies and data security measures as stated in the privacy policies.

\subsubsection{Permission Analysis}
To evaluate app permissions, we adopted a two-pronged approach. Firstly, we manually examined app permissions by installing the apps on test mobile devices. Additionally, we conducted dynamic analysis using the security assessment tool Drozer. Furthermore, we utilized the MobSF security framework to generate a comprehensive report that highlights app permissions and potential vulnerabilities that could be exploited in privacy and security attacks.

\subsubsection{Attack Surface Analysis}
When an Android app exports a component without defined permissions, app may allow malicious users and apps unauthorized access to sensitive interactions and data \cite{cwe}. To identify the vulnerable attack surface, we utilized Drozer commands to execute specific tests. Additionally, we conducted an analysis using the MobSF tool, which generated a summary of the App Components, presenting the potential attack surface.

\label{sec:results}
\section{Data Analysis and Results}
Our analyses of reproductive health tracking apps identified exploitable vulnerabilities that could allow attackers to manipulate the app's behavior, exploit identifiable device and user information, and link sensitive reproductive health data to a user's identity. These threats pose significant privacy and security risks to users, including legal prosecution. 

\subsection{Manual Analysis}
The manual analysis focuses on a thorough examination of privacy policies to identify potential exploitable flaws. Following federal guidelines on privacy protection in healthcare, it is crucial to safeguard personal identification information (PII) \cite{hhs-pii, nih-femprivacypolicy}. However, examination of the studied apps uncovered the collection of diverse identifiable data, including demographics, mobile device IDs, and individuals' reproductive health data.

The app collects user information as personal and account-related. Additionally, it collects communication/user-generated content and device information as well. The second category, known as \textit{Health Sensitive Data}, includes information such as weight, menstrual cycle data, pregnancy details (if applicable), body temperature, body measurements, symptoms, and similar health-related data.

\subsubsection{Privacy Policy: Information requested from the users}

We analyze the apps privacy policies for vulnerabilities associated with identifiable information and sensitive data collection.There are only 3 out of 20 apps offer anonymous access to the app, and majority of these apps requires an account with collecting PII and sensitive data. It is observed that 95\% of studied apps request user \textit{Email}. Other most requested identifiable data is Name (85\%), Phone Number (50\%), Address (45\%), Location (55\%), and some form of ID (15\%). Other than these primary identifiable information, age/birth year (80\%) and language (25\%) can also indicate the user demographics, while if hacked, password (45\%) can be exploited for linked attacks as users have a tendency to use exact same or similar password for multiple accounts and services \cite{riley2006password, alomari2019password}. 

Out of the studied apps, only 70\% explicitly state that they collect sensitive healthcare data. Additionally, 60\% of these apps also collect Communication/User Generated Content, while 20\% request information about others. Notably, 40\% of the apps collect pictures, and 50\% request payment information. Interestingly, 65\% of the apps declare collecting \textit{Other Information}, without disclosing specific details about the data collected under this category.
\subsubsection{Privacy Policy: Default Data Captured}
Mobile apps often collect various types of information by default, without requiring user input. It is crucial for app privacy policies to prioritize transparency, ensuring that app users are not left unaware of the information being collected by the app. Table \ref{tab:privpol-def-data} presents the types of information to collect, as stated by privacy policies of studied apps. We observe that 95\% of studied apps privacy policy declare to collect IP address. With 90\% apps using cookies, these apps are prone to HTTP cookie hijacking attacks against users on mobile devices \cite{sivakorn2016cracked}. With 85\% apps collecting \textit{Device Information}, these apps may help recovering the associated user's identity and reproductive health activities. Approximately 75\% of apps collect Time Zone/Location data, while 70\% of apps can track the features accessed by users within the app. By gathering information like network provider, browser details, usage frequency, and online activity, these apps can obtain significant data regarding users' physical and digital footprints.

\begin{table}[]
\caption{Privacy Policy: Default Data Collected }
\centering
\label{tab:privpol-def-data}
\begin{tabular}{|p{5cm}|c|}
\hline
\textbf{Default Data Collected} & \textbf{\% of Apps} \\ \hline
IP   Address                    & 95         \\ \hline
Uses Cookies                    & 90         \\ \hline
Device Information              & 85         \\ \hline
Time Zone/ Location             & 75         \\ \hline
Accessed Features Within App    & 70         \\ \hline
Internet Browser                & 65         \\ \hline
Frequency of Use                & 60         \\ \hline
Mobile Service Provider         & 35         \\ \hline
Online Activity Data            & 35         \\ \hline
Network Information             & 35         \\ \hline
\end{tabular}
\end{table}

\subsubsection{Privacy Policy: Data Handling and Data Security }
User options are vital in assessing the privacy and security of Femtech apps. However, unclear privacy policies may lead to reduced app usage and not serving as reliable resources for women in need. 

\begin{table}[!htb]
\caption{Privacy Policy: Data Handling and Security}
\centering
\label{tab:privpol-datasec}
\begin{tabular}{|p{5cm}|c|}
\hline
\textbf{Data Handling and Security}           & \textbf{\% of Apps} \\ \hline
Encrypted                            & 55         \\ \hline
Systematic Vulnerability Scanning    & 15         \\ \hline
Penetration Testing                  & 10         \\ \hline
Legal Measures                       & 15         \\ \hline
Periodic Data Protection Assessments & 15         \\ \hline
Delete / Deactivate Account          & 50         \\ \hline
Erase Data                           & 75         \\ \hline
Data Retained (Account Terminated)   & 20         \\ \hline
Data Retained (Period Not Specified) & 85         \\ \hline
Security (Not Specified/ Guaranteed) & 85         \\ \hline
\end{tabular}
\end{table}
Table \ref{tab:privpol-datasec} shows that only 10\% of apps conduct security penetration testing to identify and address vulnerabilities, compromising overall app security. 55\% of apps use data encryption for anonymization, while only 15\% perform security vulnerability scanning and periodic data protection assessments. 50\% of apps lack information on deleting or deactivating user accounts in their privacy policies. Moreover, 85\% retain user data without specifying relevant details or time periods for deleted or deactivated accounts. 20\% of apps even retain data after the account is deleted, with varying waiting periods for account deletion, such as 30 days for Clue and 7 days for the FF App. Furthermore, if users opt to share their personal information, the app may receive compensation from advertisers or sponsors in exchange for disclosing the users' data.

Additionally, one app's privacy policy states that users can request the deletion of their data; however, the app retains the option to refuse such requests and may even charge a fee if the request is deemed excessive. These uncommon or lesser-known terms and conditions may come as a surprise to app users, resulting in undesired data sharing, delays in data deletion, and potential fees to be paid.
\subsubsection{Manual Analysis of App Permissions}
During our observation of app permissions on a test mobile device, we manually collected data for both iOS and Android versions. Among iOS apps, the most commonly requested permission, requested by 50\% of the apps, is the Camera permission. On the other hand, for Android apps, the most requested permission, requested by 65\% of the apps, is the Storage permission. While iOS apps request fewer runtime permissions compared to Android apps, they do not ask for permission to access Files (20\%), Photos (35\%), Mail, and Messages (65\%). However, these apps continue to access these resources without user awareness, violating transparency and potentially compromising user privacy.

\subsection{Static Analysis}
Broadly, as categorized in Android Studio, we observed errors and warnings for four categories of \textit{Accessibility}, \textit{Correctness}, \textit{Performance}, \textit{Usability}, and \textit{Security}. We further focused on identifying major security flaws in the code. Security warnings, ranging from one to thirty-nine, indicate exploitable vulnerabilities that can be targeted for security and privacy attacks. For example, about 61\% of studied apps in Table \ref{tab:static-security-1} demonstrate that the exported service does not require permission. This indicates that these apps do not adhere to the recommended Android behavior of securing activities with minimal and necessary access \cite{androidPermissions, reardon201950}. 33\% of apps lack permissions for Content Provider, and 28.57\% of apps do not require permission for Receiver. Additionally, 52.38\% of studied apps use the dangerous feature File.setReadable() to make files world-readable, potentially compromising security. Furthermore, around 28\% of apps show the vulnerability of dynamically loading code from unsafe locations, while 23.81\% of apps carry the vulnerability of AllowBackup/FullBackupContent problems, which can be exploited through Android Debug Bridge (adb) backup and allow users to read all application data once backed up \cite{android-Studio-static, androidPermissions}.  

The utilization of private APIs by all apps can lead to crashes and data loss. These failures may result in unintended pregnancies and forced decisions, which carry legal and health risks \cite{androidPrivateAPI}.

\begin{table*}[!htb]
\centering
\caption{Static Analysis- Security Vulnerabilities}
\label{tab:static-security-1}
\begin{tabular}{|p{9.5cm}|c|p{2cm}|}
\hline
Security Vulnerabilities                                                                                                                           & \# of Apps & \%    \\ \hline
Exported service does not require permission                                                                                                       & 13         & 61.90 \\ \hline
\begin{tabular}[c]{@{}l@{}}The Network Security configuration allows \\ the use of user certificate in the release version \end{tabular} & 1          & 4.76  \\ \hline
Cipher.getInstance with ECB                                                                                                                        & 4          & 19.05 \\ \hline
File.setReadable() used to make file world-readable                                                                                                & 11         & 52.38 \\ \hline
Insecure Hostname Verifier                                                                                                                         & 4          & 19.05 \\ \hline
Receiver Does not require permission                                                                                                               & 6          & 28.57 \\ \hline
addJavascriptInterface Called                                                                                                                      & 2          & 9.52  \\ \hline
Hardware ID Uage                                                                                                                                   & 2          & 9.52  \\ \hline
Potential Multiple Cerficate Exploit                                                                                                               & 2          & 9.52  \\ \hline
Using setJavaScriptEnabled                                                                                                                         & 2          & 9.52  \\ \hline
Using the result of check permission calls                                                                                                         & 2          & 9.52  \\ \hline
load used to dynamically load code                                                                                                                 & 6          & 28.57 \\ \hline
Content provider does not require permisiion                                                                                                       & 7          & 33.33 \\ \hline
AllowBackup/FullBackupContent Problems                                                                                                             & 5          & 23.81 \\ \hline
Insecure TLS/SSL trust manager                                                                                                                     & 4          & 19.05 \\ \hline
\end{tabular}
\end{table*}

\subsection{Dynamic Analysis}
In dynamic analysis of studied reproductive health mobile apps focus on identifying dangerous permissions, attack surface, and exploitable vulnerabilities on Android apps at runtime.

App permissions analysis helps identify potential risks to users' privacy and security. Standard permissions grant access to data and actions with minimal risk, while runtime permissions provide additional access to restricted data and actions that can impact the system and other apps. Table \ref{tab:dangerous-permissions} shows the dangerous permissions captured by the tools for the studied period-tracking and fertility apps.
\subsubsection{Drozer: Dangerous Permissions}
 
\begin{table*}[!htb]
\centering
\caption{Apps Dangerous Permissions: Drozer and MobSF Analysis; 1-Identify, 0-Do not Identify}
\label{tab:dangerous-permissions}
\begin{tabular}{|p{5cm}|p{1.25cm}|p{1.25cm}|p{7cm}|}
\hline
Dangerous   Permission   & Drozer & MobSF & Exploitable Vulnerability                                                                                                                                                                           \\ \hline
ACCESS\_COARSE\_LOCATION & 1      & 1     & Malicious apps can determine approximately where user is.                                                                                                                       \\ \hline
ACCESS\_FINE\_LOCATION   & 1      & 1     & \begin{tabular}[c]{@{}l@{}}Malicious apps can determine user's location.\end{tabular}                                               \\ \hline
AUTHENTICATE\_ACCOUNTS   & 0      & 1     & \begin{tabular}[c]{@{}l@{}}Allows an app to use the account \\authenticator for creating accounts    \\ as well as obtaining and setting passwords.\end{tabular}                         \\ \hline
CALL\_PHONE              & 0      & 1     & Malicious apps may cause unexpected calls on your phone bill.                                                                                                                               \\ \hline
CAMERA                   & 1      & 1     & \begin{tabular}[c]{@{}l@{}}Allows an app to take pictures and videos \\with the camera, and collecting images \end{tabular}                            \\ \hline
GET\_ACCOUNTS            & 1      & 1     & Allows access to the list of accounts in the Accounts Service.                                                                                                                                      \\ \hline
MANAGE\_ACCOUNTS         & 0      & 1     & \begin{tabular}[c]{@{}l@{}}Allows an app to perform operations like \\adding and removing accounts    \\ and deleting their password.\end{tabular}                                             \\ \hline
POST\_NOTIFICATION       & 1      & 0     & Allows an app to post notifications                                                                                                                                                                 \\ \hline
READ\_CALENDAR           & 1      & 1     & Malicious apps can use this to send your calendar events to other people.                                                                                                                   \\ \hline
READ\_CONTACTS           & 1      & 0     & Allows an app to read the user's contacts data.                                                                                                                                             \\ \hline
READ\_EXTERNAL\_STORAGE  & 1      & 1     & Allows an app to read from external storage.                                                                                                                                                \\ \hline
READ\_PHONE\_STATE       & 1      & 1     & \begin{tabular}[c]{@{}l@{}}An app can determine the phone number \\and serial number of this phone, \\ whether a call is active, the number that call \\ is connected to and so on.\end{tabular} \\ \hline
RECORD\_AUDIO            & 1      & 1     & Allows an app to access the audio record path.                                                                                                                                                 \\ \hline
SYSTEM\_ALERT\_WINDOW    & 0      & 1     & Malicious apps can take over the entire screen of the phone.                                                                                                                                \\ \hline
USE\_CREDENTIALS         & 0      & 1     & Allows an app to request authentication tokens.                                                                                                                                             \\ \hline
WRITE\_CALENDAR          & 1      & 1     & \begin{tabular}[c]{@{}l@{}}Malicious apps can use this to delete or \\modify calendar events or to send emails.\end{tabular}                                               \\ \hline
WRITE\_EXTERNAL\_STORAGE & 1      & 1     & Allows an app to write to external storage.                                                                                                                                                 \\ \hline
\end{tabular}
\end{table*}

A per Table \ref{tab:dangerouspermissions-drozer-mobsf}, there are 12 different dangerous permission  identified in Drozer dynamic analysis \cite{drozer}. Max 62\% of apps have READ\_EXTERNAL\_STORAGE permission, while 43\% apps have WRITE\_EXTERNAL\_STORAGE. Approximately 33\% of apps have permission to track the user's approximate location through the COARSE\_LOCATION permission, while 25\% of apps can track the user's more accurate fine location, posing greater risks to privacy and security. Additionally, 29\% of apps have CAMERA permissions to access sensitive audio/video content. About 14\% of apps may have the capability to access and manipulate user accounts. Furthermore, 10\% of these apps can access audio recording functionality, and 5\% of apps have permissions to read and write the calendar.

\begin{table*}[!htb]
\centering
\caption{Dangerous Permissions Carried by Studied Period-Tracking Apps: Drozer and MobSF Analysis}
\label{tab:dangerouspermissions-drozer-mobsf}
\begin{tabular}{|p{7cm}|c|c|}
\hline
Dangerous   Permissions  & \begin{tabular}[c]{@{}c@{}}Drozer \\ (\% of   Apps)\end{tabular} & \begin{tabular}[c]{@{}c@{}}MobSF \\ (\% of Apps)\end{tabular} \\ \hline
ACCESS\_COARSE\_LOCATION & 33                                                               & 30                                                            \\ \hline
ACCESS\_FINE\_LOCATION   & 24                                                               & 20                                                            \\ \hline
AUTHENTICATE\_ACCOUNTS   & 0                                                                & 20                                                            \\ \hline
CALL\_PHONE              & 0                                                                & 5                                                             \\ \hline
CAMERA                   & 29                                                               & 25                                                            \\ \hline
GET\_ACCOUNTS            & 14                                                               & 15                                                            \\ \hline
MANAGE\_ACCOUNTS         & 0                                                                & 15                                                            \\ \hline
POST\_NOTIFICATIONS      & 29                                                               & 0                                                             \\ \hline
READ\_CALENDAR           & 5                                                                & 5                                                             \\ \hline
READ\_CONTACTS           & 5                                                                & 0                                                             \\ \hline
READ\_EXTERNAL\_STORAGE  & 62                                                               & 50                                                            \\ \hline
READ\_PHONE\_STATE       & 10                                                               & 10                                                            \\ \hline
RECORD\_AUDIO            & 10                                                               & 10                                                            \\ \hline
SYSTEM\_ALERT\_WINDOW    & 0                                                                & 5                                                             \\ \hline
USE\_CREDENTIALS         & 0                                                                & 20                                                            \\ \hline
WRITE\_CALENDAR          & 5                                                                & 5                                                             \\ \hline
WRITE\_EXTERNAL\_STORAGE & 43                                                               & 60                                                            \\ \hline
\end{tabular}
\end{table*}

\subsubsection{Drozer: Attack Surface}
Table \ref{tab:attacksurface-drozer-mobsf} shows that activities are most vulnerable with as high as 39 activities exported by a single app. We observed total 239 components exported forming the attack surface for 20 period-tracking apps. Activities exported are forming 44.77\% (107) of all exposed attack surface components identified for studied apps. Content Provide is least vulnerable with 4.18\% (10), while unprotected Service (21\% (51)) and Broadcast Receiver (29\% (71)) components may allow attackers to access sensitive information and manipulating app behavior.

\begin{table*}[!htb]
\centering
\caption{Attack Surface: Drozer and MobSF Analysis \\(AE- Activity Exported, BRE- Broadcast Receiver Exported, CPE- Content Provider Exported, SE-Service Exported)}
\label{tab:attacksurface-drozer-mobsf}
\begin{tabular}{|p{6cm}|P{.7cm}|p{.7cm}|p{.7cm}|p{.7cm}|P{.7cm}|p{.7cm}|p{.7cm}|p{.7cm}|}
\hline
                                & \multicolumn{4}{c|}{Drozer}                                                                                                                                                                                                                                                                                                                                & \multicolumn{4}{c|}{MobSF}                                                                                                                                                                                                                                                                                                                                 \\ \cline{2-9}
App Name                        & \multicolumn{1}{l|}{\begin{tabular}[c]{@{}l@{}}AE\end{tabular}} & \multicolumn{1}{l|}{\begin{tabular}[c]{@{}l@{}}BRE   \end{tabular}} & \multicolumn{1}{l|}{\begin{tabular}[c]{@{}l@{}}CPE  \end{tabular}} & \begin{tabular}[c]{@{}l@{}}SE\end{tabular} & \multicolumn{1}{l|}{\begin{tabular}[c]{@{}l@{}}AE\end{tabular}} & \multicolumn{1}{l|}{\begin{tabular}[c]{@{}l@{}}BRE   \end{tabular}} & \multicolumn{1}{l|}{\begin{tabular}[c]{@{}l@{}}CPE   \end{tabular}} & \begin{tabular}[c]{@{}l@{}}SE\end{tabular} \\ \hline
Always You: Period Tracker      & \multicolumn{1}{l|}{2}                                                                & \multicolumn{1}{l|}{3}                                                                           & \multicolumn{1}{l|}{1}                                                                         & 2                                                              & \multicolumn{1}{l|}{1}                                                                & \multicolumn{1}{l|}{3}                                                                           & \multicolumn{1}{l|}{1}                                                                         & 3                                                              \\ \hline
Birth Control - Natural Cycles  & \multicolumn{1}{l|}{2}                                                                & \multicolumn{1}{l|}{2}                                                                           & \multicolumn{1}{l|}{0}                                                                         & 1                                                              & \multicolumn{1}{l|}{3}                                                                & \multicolumn{1}{l|}{3}                                                                           & \multicolumn{1}{l|}{1}                                                                         & 2                                                              \\ \hline
Clover - Safe Period Tracker    & \multicolumn{1}{l|}{1}                                                                & \multicolumn{1}{l|}{2}                                                                           & \multicolumn{1}{l|}{0}                                                                         & 1                                                              & \multicolumn{1}{l|}{0}                                                                & \multicolumn{1}{l|}{1}                                                                           & \multicolumn{1}{l|}{0}                                                                         & 2                                                              \\ \hline
Clue - Period \& Cycle Tracker  & \multicolumn{1}{l|}{39}                                                               & \multicolumn{1}{l|}{4}                                                                           & \multicolumn{1}{l|}{0}                                                                         & 5                                                              & \multicolumn{1}{l|}{38}                                                               & \multicolumn{1}{l|}{4}                                                                           & \multicolumn{1}{l|}{0}                                                                         & 5                                                              \\ \hline
Eve - Period Tracker            & \multicolumn{1}{l|}{7}                                                                & \multicolumn{1}{l|}{7}                                                                           & \multicolumn{1}{l|}{4}                                                                         & 5                                                              & \multicolumn{1}{l|}{6}                                                                & \multicolumn{1}{l|}{7}                                                                           & \multicolumn{1}{l|}{4}                                                                         & 6                                                              \\ \hline
Fertility Friend FF App         & \multicolumn{1}{l|}{2}                                                                & \multicolumn{1}{l|}{1}                                                                           & \multicolumn{1}{l|}{0}                                                                         & 0                                                              & \multicolumn{1}{l|}{1}                                                                & \multicolumn{1}{l|}{1}                                                                           & \multicolumn{1}{l|}{0}                                                                         & 1                                                              \\ \hline
Flo                             & \multicolumn{1}{l|}{7}                                                                & \multicolumn{1}{l|}{4}                                                                           & \multicolumn{1}{l|}{0}                                                                         & 3                                                              & \multicolumn{1}{l|}{6}                                                                & \multicolumn{1}{l|}{4}                                                                           & \multicolumn{1}{l|}{0}                                                                         & 4                                                              \\ \hline
Glow                            & \multicolumn{1}{l|}{8}                                                                & \multicolumn{1}{l|}{7}                                                                           & \multicolumn{1}{l|}{3}                                                                         & 5                                                              & \multicolumn{1}{l|}{7}                                                                & \multicolumn{1}{l|}{7}                                                                           & \multicolumn{1}{l|}{3}                                                                         & 6                                                              \\ \hline
Kindara: Fertility Tracker      & \multicolumn{1}{l|}{2}                                                                & \multicolumn{1}{l|}{1}                                                                           & \multicolumn{1}{l|}{0}                                                                         & 4                                                              & \multicolumn{1}{l|}{1}                                                                & \multicolumn{1}{l|}{1}                                                                           & \multicolumn{1}{l|}{0}                                                                         & 4                                                              \\ \hline
Luna                            & \multicolumn{1}{l|}{1}                                                                & \multicolumn{1}{l|}{1}                                                                           & \multicolumn{1}{l|}{0}                                                                         & 0                                                              & \multicolumn{1}{l|}{0}                                                                & \multicolumn{1}{l|}{1}                                                                           & \multicolumn{1}{l|}{0}                                                                         & 0                                                              \\ \hline
MagicGirl/Teen   Period Tracker & \multicolumn{1}{l|}{1}                                                                & \multicolumn{1}{l|}{1}                                                                           & \multicolumn{1}{l|}{0}                                                                         & 2                                                              & \multicolumn{1}{l|}{0}                                                                & \multicolumn{1}{l|}{1}                                                                           & \multicolumn{1}{l|}{0}                                                                         & 2                                                              \\ \hline
Maya                            & \multicolumn{1}{l|}{5}                                                                & \multicolumn{1}{l|}{5}                                                                           & \multicolumn{1}{l|}{0}                                                                         & 2                                                              & \multicolumn{1}{l|}{4}                                                                & \multicolumn{1}{l|}{5}                                                                           & \multicolumn{1}{l|}{0}                                                                         & 2                                                              \\ \hline
My Calendar                     & \multicolumn{1}{l|}{1}                                                                & \multicolumn{1}{l|}{6}                                                                           & \multicolumn{1}{l|}{0}                                                                         & 4                                                              & \multicolumn{1}{l|}{0}                                                                & \multicolumn{1}{l|}{6}                                                                           & \multicolumn{1}{l|}{0}                                                                         & 4                                                              \\ \hline
MyDays X                        & \multicolumn{1}{l|}{1}                                                                & \multicolumn{1}{l|}{2}                                                                           & \multicolumn{1}{l|}{1}                                                                         & 3                                                              & \multicolumn{1}{l|}{0}                                                                & \multicolumn{1}{l|}{2}                                                                           & \multicolumn{1}{l|}{1}                                                                         & 3                                                              \\ \hline
Ovia Fertility \& Cycle Tracker & \multicolumn{1}{l|}{5}                                                                & \multicolumn{1}{l|}{6}                                                                           & \multicolumn{1}{l|}{0}                                                                         & 2                                                              & \multicolumn{1}{l|}{4}                                                                & \multicolumn{1}{l|}{6}                                                                           & \multicolumn{1}{l|}{0}                                                                         & 3                                                              \\ \hline
P.Tracker                       & \multicolumn{1}{l|}{2}                                                                & \multicolumn{1}{l|}{2}                                                                           & \multicolumn{1}{l|}{1}                                                                         & 4                                                              & \multicolumn{1}{l|}{1}                                                                & \multicolumn{1}{l|}{2}                                                                           & \multicolumn{1}{l|}{1}                                                                         & 4                                                              \\ \hline
Period Diary                    & \multicolumn{1}{l|}{4}                                                                & \multicolumn{1}{l|}{1}                                                                           & \multicolumn{1}{l|}{0}                                                                         & 2                                                              & \multicolumn{1}{l|}{3}                                                                & \multicolumn{1}{l|}{1}                                                                           & \multicolumn{1}{l|}{0}                                                                         & 3                                                              \\ \hline
Premom Ovulation Tracker        & \multicolumn{1}{l|}{8}                                                                & \multicolumn{1}{l|}{5}                                                                           & \multicolumn{1}{l|}{0}                                                                         & 2                                                              & \multicolumn{1}{l|}{-}                                                                 & \multicolumn{1}{l|}{-}                                                                            & \multicolumn{1}{l|}{-}                                                                          & -                                                               \\ \hline
Spot On Period Tracker          & \multicolumn{1}{l|}{2}                                                                & \multicolumn{1}{l|}{4}                                                                           & \multicolumn{1}{l|}{0}                                                                         & 2                                                              & \multicolumn{1}{l|}{1}                                                                & \multicolumn{1}{l|}{4}                                                                           & \multicolumn{1}{l|}{0}                                                                         & 3                                                              \\ \hline
Stardust Period Tracker         & \multicolumn{1}{l|}{7}                                                                & \multicolumn{1}{l|}{7}                                                                           & \multicolumn{1}{l|}{0}                                                                         & 2                                                              & \multicolumn{1}{l|}{10}                                                               & \multicolumn{1}{l|}{4}                                                                           & \multicolumn{1}{l|}{1}                                                                         & 2                                                              \\ \hline
\end{tabular}
\end{table*}

\subsubsection{Drozer: Injection Vulnerabilities}
SQL injection attack can expose private data, corrupt database contents, and even compromising of back-end infrastructure \cite{android-injection}. Among the studied apps, 19 out of 20 did not show any SQL injection vulnerabilities in content providers. However, one app was found to be vulnerable, with instances of projection and selection vulnerabilities. Injection vulnerabilities can lead to the exposure of sensitive data, bypassing security measures, and potential damage to databases. The impact includes risks to user privacy, loss of intellectual property, and compromised trust in app providers \cite{android-injection}.

\subsection{MobSF Analysis}
We performed the MobSF analysis to compare the app privacy and security vulnerabilities compared with other analysis performed in this study. MobSF provides an overall security score (1-100) to show the security risk level of the app. 10\% apps show \textit{Low Risk}, 75\% apps are with \textit{Medium Risk} with 40-59 score, and other 15\% apps show \textit{High Risk}.

\noindent{\textbf{MobSF- Dangerous Permissions:}}
A per Table \ref{tab:dangerouspermissions-drozer-mobsf}, there are 15 different dangerous permission  identified MobSF analysis. Max 50\% of apps have READ\_EXTERNAL\_STORAGE permission, while 60\% apps have WRITE\_EXTERNAL\_STORAGE. Coarse location permission can track the user's approximate location and 30\% apps have permission to do so, while 20\% apps can track fine location of users which is potentially more damaging to user's privacy and security. 25\% apps have Camera permissions to access sensitive audio/video content. 20\% of apps show permission to authenticate account. 15\% of Apps may have capability to access and manipulate the accounts. 10\% of these apps can access audio record functionality and 5\% of apps can read and write the calendar.

\noindent{\textbf{MobSF- Attack Surface:}}
There are a total of 220 components exported identified in MobSF analysis. There are are 39\% (86) Activity Exported (AE), Broadcast Receiver Exported (BRE) as 29\% (63), Content Provider Exported (CPE) as 5\%(12), and 27\% (59) are Service Exported (SE).

\noindent{\textbf{MobSF- Network Security: }} According to MobSF analysis, apps are considered highly vulnerable if their Domain config allows clear text traffic to domains within the scope. 

\noindent{\textbf{MobSF- Code Analysis: }} 
Code analysis by MobSF identifies the vulnerabilities in app code.apk files with the security standards defined by OWASP and CWE \cite{owasp-masvs, cwe}. There are eighteen code vulnerabilities detected, and eleven out of eighteen issues fall into OWASP Top 10 vulnerabilities. 
 
20\% of apps show a high risk for insecure communication, inadequate integrity checking, and improper platform usage. These vulnerabilities can lead to Man-in-the-Middle (MITM) attacks and Padding Oracle attacks, which can result in unauthorized access to data and decryption of protected sensitive reproductive health information. Other concerning code vulnerabilities include the usage of hardware IDs, disclosure of IP addresses, SQL injection, and hard-coded sensitive information.

\noindent{\textbf{MobSF- Network Security and Certificate Analysis: }} 
20\% of studied apps found as \textit{Secure} on network security, and 15\% of apps present high risk. Insecure configuration refers to settings that do not enforce secure communication protocols, such as using encryption or secure channels for transmitting sensitive data. Clear text traffic refers to data transmitted without encryption, making it susceptible to interception and unauthorized access. About 60\% of studied apps do not show information about network security. 

Android apps are signed with digital certificates to ensure authenticity and integrity, with multiple signatures (v1, v2, and v3) available to maintain validity and security. A false signature could indicate that the app's certificate has been tampered with or modified, potentially indicating malicious intent. MobSF analysis shows that 20\% of apps have v1 signature as false, and 35\% of studied period-tracking apps show v3 signature as false. A violation of an app's authenticity and integrity can call into question the validity of its behavior and guidance for managing monthly cycles or fertility monitoring. 

\noindent\textbf{Threat of Hidden and Undetected Dangerous Permissions:} In our manual analysis, we identified seven different types of dangerous permissions. However, MobSF identified fifteen dangerous permissions, compared to twelve identified by Drozer. Table \ref{tab:dangerous-permissions} provides a summary of all the dangerous permissions observed by both tools. Interestingly, Drozer revealed a larger attack surface (239) compared to MobSF (220), indicating a higher potential for security vulnerabilities. Different tools may detect and handle app vulnerabilities differently, leading to varying results in vulnerability scans. Additionally, our research has revealed that users may lack complete information about dangerous permissions requested by apps. For instance, users may not know whether an app is tracking their precise (FINE) or approximate (COARSE) location or both when granting "Location" permission. This lack of transparency raises privacy concerns and highlights the importance of improved communication between app developers and users regarding the use of dangerous permissions.

\subsection{Domain-Specific Implications}
In the context of period tracking and pregnancy monitoring apps, these privacy and security breaches can be particularly alarming. Users may be sharing highly sensitive and personal information about their menstrual cycles, fertility, and reproductive health. If this information is mishandled, it can have serious consequences, such as social stigma, discrimination, or even the possibility of being used against the user in legal or personal matters.

If permissions are absent, bypassed or abused, it can result in various privacy and security breaches. It raises concerns about potential tracking or surveillance, particularly troubling in the context of pregnancy or abortion, where individuals may seek to keep their reproductive choices private and confidential. Storing and sharing of unique device identifiers could lead to the tracking and profiling of women trying to conceive or avoiding pregnancy. Location data can reveal sensitive information about a user's daily routines, potentially compromising their safety and even restricting them to get needed medical assistance.

These apps may store sensitive information, such as medical records, ultrasound images, or personal notes related to the period or pregnancy. Pregnant women may share sensitive information within these apps, including health conditions, prenatal care details, or even personal thoughts and emotions. A breach to storage data can result in various risks, such as identity theft, blackmail, or other forms of exploitation. Moreover, access to contacts can be misused for unsolicited marketing or even social engineering attacks. Invasive permissions like camera and microphone access can invade  privacy, capturing personal and intimate moments without their knowledge or consent.

As observed in our study, the presence of an injection vulnerability in a Femtech app pose severe consequences, such as unauthorized code execution, data tampering, and privacy breaches. Moreover, injection vulnerabilities can also enable attackers to modify or manipulate the app's functionalities, leading to unexpected or incorrect behavior. In the context of women's health, this could result in inaccurate ovulating period, due date calculations, incorrect medication reminders, or other critical errors that may impact the user's health and well-being. This may lead to unintended pregnancies and potentially forcing individuals into life-threatening health situations if safe abortion options are unavailable.

Exploiting calendar access or manipulating notifications in pregnancy tracking apps can lead to misinterpretation of ovulation cycles, increasing the risk of unintended pregnancies. This is particularly concerning considering the existing burden of teen pregnancies as a social and health issue. Additionally, fine-location tracking features in these apps can pose a threat of physical attacks, especially for individuals seeking abortions for personal, social, financial, or health reasons. This can result in heightened fear among women who may be concerned about being tracked and facing unwarranted legal repercussions.

\label{sec:Limitations and Challenges}
\section{Challenges and Limitations}

The app selection process was limited to find a large number of apps which are supported by iOS and Android both. To keep the study more practical, we analyzed free apps with better accessibility to intended population, which also limited some of the analysis for certain apps included in the study. For example, MobSF could not analyze the apps with very large .apk files. We collected some data and then reshuffled the apps list to keep it consistent with the manual, dynamic, and static analysis by utilized tools. On vulnerabilities, we observed that there are many privacy and security warnings observed in the code analysis, but it is out of scope to explain every vulnerability which can be exploited to manipulate the app and user behavior resulting in unintended outcomes. This study is limited in parts to examine both iOS and Android app counterparts. For example, Android apps are more in-depth analyzed with the availability of .apk files. 

Period tracking and fertility monitoring apps indeed undergo frequent updates and additions of new features. This dynamic nature poses a challenge in identifying vulnerabilities as security assessments need to keep up with the constantly evolving app versions. Additionally, app statistics, such as user base, ratings, and number of reviewers, are also subject to change over time, making it important to consider the most up-to-date information when evaluating the app's reputation and user feedback. Scope of this work is limited to analyze unpaid versions. Additionally, it is important to acknowledge that vulnerability testing tools  may not provide an exhaustive assessment. They can help identify common vulnerabilities but may not uncover all possible security weaknesses. \bluetext{While our focus in this study does not encompass the actual implementation of attacks, it serves as a foundational exploration of the threats that could potentially exploit FemTech apps. }

\label{sec:discussion and future work}

\section{Further Discussion}

Period-tracking apps are fast growing to assist with reproductive healthcare needs of women, changing healthcare practices, and social environment. Legal disparity exacerbates the contextual threat posed by vulnerabilities in reproductive health mobile apps. These privacy and security threats have a far-reaching impact on women's reproductive health, affecting them physically, mentally, and emotionally. However, the consequences are even more severe due to the criminalization of abortion, as women bear a significantly greater burden compared to their male counterparts. This disparity in accountability further compounds the challenges faced by women, intensifying the overall impact on their reproductive health.

Amid the rapid development of pregnancy monitoring mobile apps to meet market needs, it is crucial to prioritize addressing general vulnerabilities, such as security and data protection, while also ensuring usability and accessibility considerations are given due attention. Neglecting these aspects can lead to compromised user experiences, potential misuse of the app's functionalities, and exclusion of users with diverse needs, ultimately hindering the app's effectiveness and broad adoption. Developers have a responsibility to minimize the use of dangerous permissions in pregnancy monitoring mobile apps. Unnecessary access to sensitive features or data should be avoided, reducing potential privacy and security risks. Additionally, when users deny certain permissions, developers should handle these denials appropriately, gracefully degrading the app's functionality without compromising usability. Enforced guidelines and rules can make developers more accountable while following best practices proposed by development frameworks and the federal government can minimize attack surfaces. Authentic guidance for selecting reproductive health management apps, backed by clinical trials and vetted by authorities such as FDA, FTC, NIH, and HHS, can help users make secure and safe choices.

Post Roe v. Wade, there is an increased attention by app developers to provide anonymous access but implementation may take longer with technical and operational challenges \cite{flo-anonymous, flo-anonymous-whitepaper}. We should not be waiting for major incidents to happen before being forced to react. Developing a framework for evaluating period-tracking apps pre-market may minimize the post-market privacy and security breaches. It can also reduce the burden on legal system and the cost of unintended pregnancies, which is still a known healthcare burden in the United States \cite{stevenson2021pregnancy, cdc-teen-preg-2019, cdc-abortion-2020}. A longitudinal study can observe patterns in developing sensitive health data handling apps with improved security by design, informing future policies to protect user privacy and security for women's health apps. 

\label{sec:conclusion}
\section{Conclusion}

In this work, we analyze the privacy and security vulnerabilities of period-tracking mobile apps impacting women reproductive health management. Availability of these apps are promising to provide a way to reach out to disadvantaged population, which is at greater risk without proper needed medical consultation and assistance. However, these apps can be exploited in multiple ways because of existing technical loopholes. App development faces challenges in complying with secure practices due to rapidly changing market needs. Period-tracking and fertility apps lack HIPAA regulation but are subject to FTC rules for protecting sensitive health data. Transparent privacy policies help users make informed decisions and trust these apps. However, relying solely on peer evaluations may overlook privacy and security pitfalls in the apps' policies.
 
Manipulated period-tracking apps can result in unintended pregnancies and risky decisions, such as abortion. Identifying and tracking users' reproductive patterns post Roe v. Wade poses a serious threat to an already burdened population. Inaccessible medical resources further endanger patient health and lives during illegal abortions \cite{stevenson2021pregnancy}.

Our study exposed privacy and security vulnerabilities in period-tracking and fertility-monitoring apps, posing legal risks in the post Roe v. Wade era. These threats strain social, financial, and legal resources. Defining and enforcing standards can ensure app developers adhere to security-by-design principles and accountability. Collaboration among medical professionals, users, developers, and legal experts fosters secure digital health management tools, reducing the overall healthcare burden.


\newcommand{\BIBdecl}{\setlength{\itemsep}{0.25 em}}
\raggedright
\bibliographystyle{plain}

\bibliography{main}

\end{document}